\documentclass[preprint,5p]{elsarticle}
\usepackage{natbib}
\usepackage{graphics}
\usepackage{subfigure}
\usepackage{nicefrac}

\usepackage{amssymb}
\usepackage{amsthm}
\usepackage{siunitx}

\biboptions{sort&compress}

\journal{Nuclear Instruments and Methods in Physics Research A }

\begin{document}

\begin{frontmatter}



\title{$\beta$-particle energy-summing
correction for $\beta$-delayed proton emission measurements}


\author[Add1,Add2]{Z.~Meisel\corref{cor1}}
\ead{meisel@ohio.edu}
\address[Add1]{Institute of Nuclear and Particle Physics, Department of Physics and Astronomy, Ohio University, Athens, OH 45701, USA}
\address[Add2]{Joint Institute for Nuclear Astrophysics -- Center
for the Evolution of the Elements, www.jinaweb.org}
\address[Add3]{National Superconducting Cyclotron Laboratory, Michigan State University, East Lansing, MI 48824, USA}
\address[Add4]{Nuclear Science Division, Lawrence Berkeley National Laboratory, Berkeley, CA 94720, USA}
\address[Add5]{Grand Acc\'{e}l\'{e}rateur National d'Ions Lourds, CEA/DSM-CNRS/IN2P3, Caen 14076, France}
\address[Add6]{Institute for Applied Physics, Goethe University Frankfurt am Main, 60438 Frankfurt am Main, Germany}
\address[Add7]{Department of Physics and Astronomy, Michigan State University, East Lansing, MI 48824, USA}
\address[Add8]{Department of Physics and
Astronomy, University of Tennessee, Knoxville, TN 37996, USA}
\cortext[cor1]{Corresponding author}

\author[Add2,Add3]{M.~del~Santo}

\author[Add4]{H.L.~Crawford}

\author[Add2,Add3]{R.H.~Cyburt}

\author[Add5]{G.F.~Grinyer}

\author[Add2,Add6]{C.~Langer}

\author[Add2,Add3]{F.~Montes}

\author[Add2,Add3,Add7]{H.~Schatz}

\author[Add2,Add8]{K.~Smith}

\address{}

\begin{abstract}

A common approach to
studying $\beta$-delayed proton emission is to measure the energy of
the emitted proton and corresponding nuclear recoil 
in a double-sided silicon-strip detector (DSSD)
after implanting the $\beta$-delayed proton-emitting
($\beta$p) nucleus.  However, in order to extract the proton-decay
energy,
the measured
energy must be
corrected for the additional energy implanted in the DSSD by the
$\beta$-particle emitted from the $\beta$p nucleus, an effect
referred to here as $\beta$-summing.  We present an approach to
determine an accurate correction for $\beta$-summing. Our method
relies on the determination of the
mean implantation depth of the $\beta$p nucleus within the DSSD by
analyzing the shape of the total (proton + recoil + $\beta$) decay energy
distribution shape. We validate this approach with
other mean implantation depth measurement techniques that take advantage
of energy deposition within DSSDs upstream and downstream of the
implantation DSSD.

\end{abstract}


\end{frontmatter}


\section{Introduction}
\label{Introduction}

$\beta$-delayed proton emission experiments can be used to
populate proton-emitting states in nuclei which are otherwise
difficult to access.  Such proton-emitting states can be of interest
to nuclear astrophysics and
nuclear structure studies~\cite{DSan14,Loru12,Borg13}.
Rare-isotope beam facilities that produce short-lived isotopes
either by projectile fragmentation or using the isotope separation
online (ISOL) technique provide
the opportunity to study the exotic nuclei which exhibit $\beta$-delayed proton emission, $\beta$p nuclei.  

To date, the most common method for studying $\beta$p nuclei employs
the production of nuclei via projectile fragmentation followed by
the subsequent implantation of the $\beta$p
nucleus into a double-sided silicon-strip detector (DSSD) with a
thickness on the order of 50-1000~$\mu$m 
\cite[e.g. References][]{DSan14,Loru12,Saas11,Orri14,Piec95,Blan97}.  
After implantation into the DSSD, the $\beta$p nucleus undergoes
$\beta$-delayed proton emission and the energy of the proton and the
corresponding nuclear recoil (referred to here as the proton-decay
energy) is then
detected within the DSSD.  The high degree of segmentation in the
DSSD, in the case studied here forty 1~mm-pitch strips in both planar dimensions, enables the correlation of implants and decays, even with
implantation rates on the order of 1~kHz.  This capability has
the advantage of turning the contaminant nuclei in the beam
whose decays are
well understood into calibration nuclei.  

One consequence of using
the projectile fragmentation production method is a relatively large
energy spread of the ions of interest that impinge on the
implantation DSSD.  This large energy spread
requires a relatively thick DSSD to
stop all ions of interest in the active area of the detector.  However, an undesired consequence of
this approach is the additional energy deposited in the DSSD from
the $\beta$-particle which is emitted almost simultaneously with the
proton in $\beta$-delayed proton emission.  Therefore what is
measured is not the proton-decay energy, but rather the total decay
energy. This is a phenomenon we
refer to here as the $\beta$-summing effect. An example in the literature
is shown in
Figure 8 of Reference~\cite{Saas11}, where it is evident that the Gaussian
peak of proton energy deposition is shifted to higher energy and
convolved with a high-energy tail due to the addition of
$\beta$-particle energy deposition. The $\beta$-summing effect is one of the
dominant uncertainties in determining the energy deposited by a
proton decay within
a DSSD after a $\beta$-delayed proton emission event, contributing
several tens of keV to energy uncertainties of $\sim$50--100~keV for
proton-decay
energies of a few MeV~\cite{DSan14,Orri14}. As such, it is necessary
to take into account $\beta$-summing for studies of $\beta$p nuclei
whose results rely on a precise proton-decay energy determination, which is often
the case for nuclear astrophysics~\cite[e.g. Reference][]{DSan14}.

We present here an approach to addressing the problem of
$\beta$-summing.  This method, which was employed in the analysis of
data presented in Reference~\cite{DSan14}, consists of determining the
mean implantation depth of a $\beta$p nucleus within a DSSD by
reproducing the measured shape of the total (proton + recoil + $\beta$)
energy-deposition histogram for
$\beta$-delayed proton-emission events with simulations.  The following
section, Section~\ref{MeasAndSim}, 
discusses the data collected and {\tt GEANT4}~\cite{Agos03} simulations used to accomplish the
$\beta$-summing correction analysis, as well as the simulation
validation. Section~\ref{Depth} presents our newly developed mean
implantation depth determination method, which is essential in
determining the $\beta$-summing correction, and Section~\ref{Compare}
provides comparisons to alternative mean implantation depth
determination methods. Section~\ref{Conclusions} describes the
full process of obtaining the $\beta$-summing correction.

\begin{figure}[ht]\begin{center}
\includegraphics[width=1.0\columnwidth,angle=0]{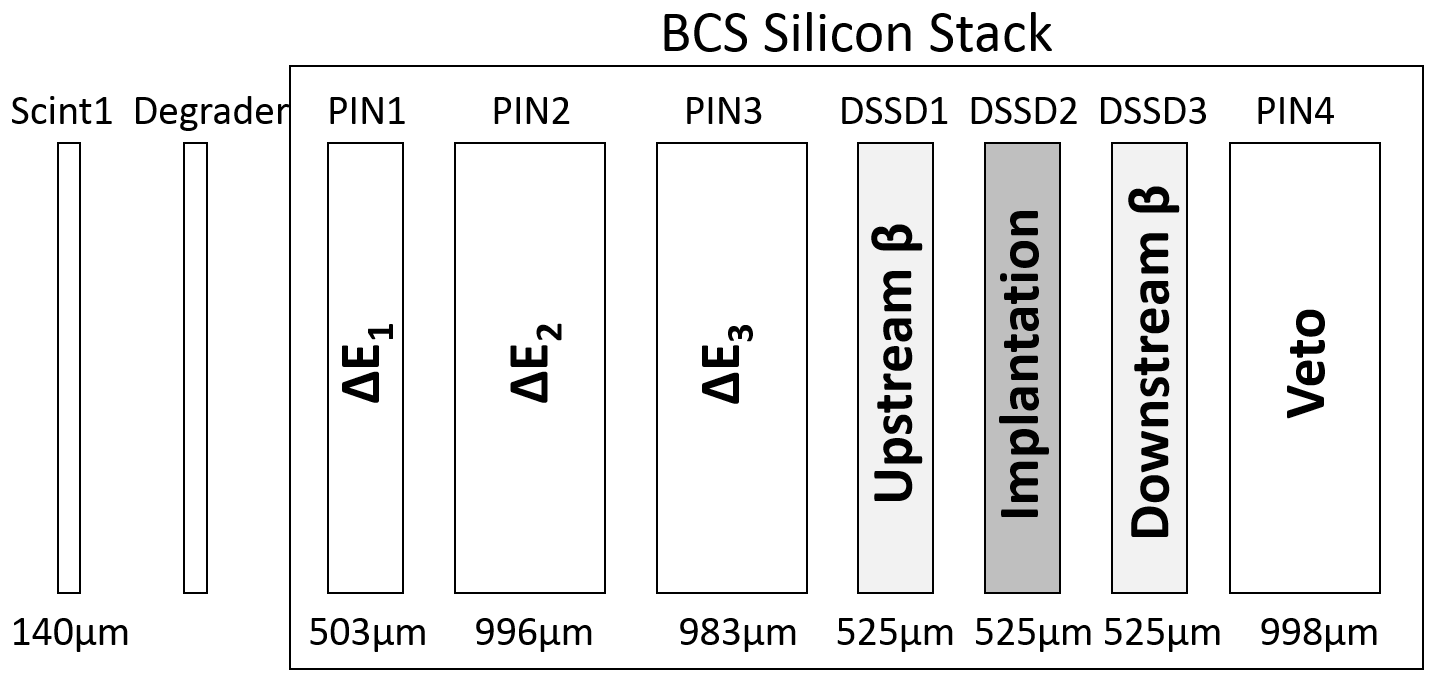}
\caption{Schematic arrangement of the silicon detectors within the
NSCL BCS in the configuration used for the experimental campaign
described in the text.
\textit{Scint1} is a BC400 timing scintillator, \textit{Degrader} is an aluminum foil, \textit{PIN}
are single-sided silicon strip detectors, and
\textit{DSSD} are double-sided silicon-strip
detectors. The thickness of each detector is indicated below it in
the figure and the role of each detector is indicated by the label
within the detector.
Relative longitudinal spacings of the detectors are listed in Table~\ref{DetectorTable}.}
\label{BCSdraw}
\end{center}\end{figure}

\begin{center}
\begin{table}
  \caption{Thicknesses and relative longitudinal positions of the
  detectors' central planes (along the
  beam axis) with respect to DSSD2, the detector in which
  $\beta$p nuclei were implanted.  See Figure~\ref{BCSdraw} for a
  schematic representation.}
\centering
  \begin{tabular}{ c  c  c }
  \hline
  Detector &  Thickness [mm] & Relative position [mm] \\ \hline
  PIN1  & 0.503 & -7.73 \\ 
  PIN2  & 0.996 & -5.87 \\ 
  PIN3  & 0.983 & -3.68 \\ 
  DSSD1 & 0.525 & -1.73 \\ 
  DSSD2 & 0.525 &  0.00 \\ 
  DSSD3 & 0.525 &  2.23 \\ 
  PIN4  & 0.998 &  4.19 \\ 
  \end{tabular}
  \label{DetectorTable}
  \end{table}
\end{center}

\section{Measurements and Simulations}
\label{MeasAndSim}
We focus on the $\beta$-summing correction which was developed using
data from a $\beta$-delayed proton emission experimental campaign
performed at the National Superconducting Cyclotron Laboratory
(NSCL),
partially described in Reference~\cite{DSan14}.  The data chosen are typical
of that collected on $\beta$p nuclei produced via
projectile fragmentation, with subsequent implantation into a
relatively thick DSSD.
Additionally, properties of the studied nuclei are well known from
previously published data. As such, they provide an ideal case to
study our proposed $\beta$-summing correction method.
{\tt GEANT4}~\cite{Agos03} was
chosen to simulate the measurements due to its flexibility and
rigorously validated physics packages~\cite[e.g. Reference][]{Bati13}. The following subsections provide
detailed descriptions of the experimental data collection and the
simulations thereof.

\subsection{$\beta$-summing data collection}
\label{DataForSimVal}
The data to assess $\beta$-summing were collected using $\beta$p emitting
nuclei which were produced via projectile fragmentation using the Coupled
Cyclotron Facility at the NSCL~\cite{York99}.  The primary beam
used to produce the $\beta$p
nuclei $^{23}$Si and $^{20}$Mg was $^{36}$Ar, whereas a $^{78}$Kr
primary beam was used to produce the $\beta$p nucleus $^{69}$Kr.
The primary beams were impinged on a beryllium target of
thickness 1060~mg/cm$^{2}$ for $^{36}$Ar and 141~mg/cm$^{2}$ for
$^{78}$Kr at an energy of
150~MeV/u. The resultant cocktail beams were purified first with the
A1900 fragment separator~\cite{Morr03} and then further purified via
the Radio Frequency Fragment Separator~\cite{Bazi09} prior to
implantation within the central DSSD of the Beta Counting System
(BCS)~\cite{Pris03}, where $\beta$-delayed proton emission events
were measured. The implantation rate within the BCS was
$\sim$100~Hz.

A schematic of the experiment set-up is given in
Figure~\ref{BCSdraw} and the relative longitudinal positions of detectors within
the BCS are listed in Table~\ref{DetectorTable}. Particle identification of implanted nuclei
was performed using the $\Delta$E--TOF technique, where timing signals
were provided by the Coupled Cyclotron RF and a
140~$\mu$m-thick BC400 scintillator upstream of the BCS and three PIN
detectors, of thickness 
503~$\mu$m, 996~$\mu$m, and 983~$\mu$m. The PIN detectors were also
used to measure energy loss prior to the implantation DSSD.  An aluminum
degrader upstream of the BCS was used to ensure implantation of
nuclei of interest within the central DSSD of the BCS by adjusting
the degrader's angle with respect to the beam direction. A stack of
three 
525~$\mu$m
DSSDs, model BB1 from Micron Semiconductor Ltd., located
downstream of the first three PIN detectors in the BCS were used to
detect implantations from beam fragments and $\beta$-particles from
subsequent decays. The central DSSD was intended for implantation of
$\beta$p nuclei. A single 
998~$\mu$m-thick PIN detector, located downstream of
the DSSD stack within the BCS, was used to veto light beam fragments
that could be mistaken for $\beta$-particles by rejecting events
with a coincident upstream PIN detection. 
Each DSSD is segmented into
40$\times$40 strips
with a 1~mm pitch, i.e. consisting of 1600 1$\times$1~mm$^{2}$ virtual pixels.
Each PIN detector is 5$\times$5~cm$^{2}$ and has no segmentation.
The detectors within the
BCS were surrounded by the Segmented Germanium Array (SeGA) in the
beta-configuration~\cite{Muel01} in order to measure proton emission
branchings and decays to non-proton-emitting states with high
resolution.

$\beta$-delayed
proton emission events were correlated in time with 
$\beta$p nuclei previously implanted in the same DSSD pixel, while
SeGA provided complementary $\gamma$-detection.
A notable contaminant which accompanied the $\beta$p nucleus
$^{69}$Kr was $^{67}$Se (See Figure 2 of Reference~\cite{DSan14}.), as this
was used to determine the implantation DSSD
detection threshold (See Section~\ref{G4simVal}.).
Before and after the projectile fragmentation experiments, the
calibration $\alpha$-source $^{228}$Th was employed to characterize the
energy resolution of the DSSDs.

\subsection{{\tt GEANT4} simulations}
\label{G4sim}
We employ the Monte Carlo particle transport software
{\tt GEANT4}~\cite{Agos03} version 9.6.02 to simulate the
$\beta$-delayed proton emission measurements described in the
previous subsection.
Detector features such as geometry, orientation, and resolution were
included in the {\tt GEANT4} simulations, though for the purpose of
simplification the BCS was modeled as a stack of free-floating
silicon detectors within an aluminum cylinder.


$\beta$-energy spectra were sampled from the $\beta$-decay
distribution given by the Fermi theory of
$\beta$-decay~\cite{Ferm34}, defined by the $Q$-value between the
initial and final states. 
Strictly speaking, the final shape of the
$\beta$-decay energy distribution depends on nuclear corrections
related to the charge and size of the parent
nucleus~\cite{Rose36,Sirl67,Sirl87,Wilk89,Wilk90}. However, these
higher-order nuclear corrections result in a relatively minor change
(e.g. compared to small changes in the decay $Q$-value) in the
overall decay spectrum shape~\cite{Venk85,Geor14} and so these
nuclear corrections can be safely neglected. Though the $Q$-value
significantly affects the energy spectrum of emitted
$\beta$-particles, we demonstrate (in Section~\ref{Qsens}) a relatively
small sensitivity of our reported results to the choice of
$\beta$-decay $Q$-value, which could vary due to mass uncertainties
and/or different assumed final states of the decay.

\subsection{Simulation validation}
\label{G4simVal}

Since a proton from an implanted $\beta$p nucleus with an energy
up to a few MeV has a mean
free path up to tens of
microns\footnote{http://physics.nist.gov/Star}, it generally
deposits 100\% of its energy within
the DSSD, as does the corresponding nuclear recoil, and therefore the critical element of the {\tt GEANT4} simulations requiring
validation is the partial energy deposition of the $\beta$-particle within
the DSSD.
The simulation of $\beta$-particle energy deposition within the DSSD was validated via comparison to
experimental measurements of the well-studied $\beta$-emitter
$^{67}$Se~\cite{Jund05}, which accompanied the production of $^{69}$Kr as a
contaminant from fragmentation of $^{78}$Kr (Shown in Figure~2
of Reference~\cite{DSan14}.).

\begin{figure}[ht]\begin{center}
\includegraphics[width=1.0\columnwidth,angle=0]{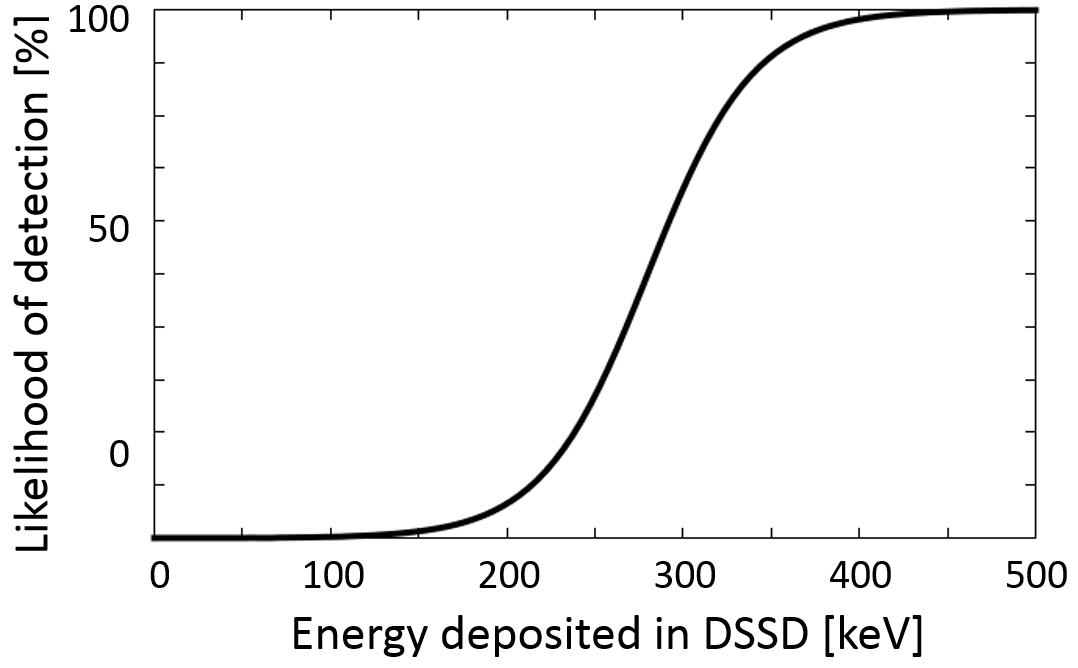}
\caption{Likelihood of event detection as a function of
energy deposited within the implantation DSSD, as given by
Equation~\ref{threshEqn} with $c$=60~keV and $d$=280~keV.
 }
\label{ThreshFn}
\end{center}\end{figure}

Comparison to data required a determination of the energy resolution
and energy detection threshold of the implantation DSSD.  The
150~keV FWHM energy
resolution, which was observed to be dominated by electronic noise
and thus independent of energy,
was determined by fitting several Gaussian distributions to the
spectrum of a $^{228}$Th $\alpha$-source.
We used a probabilistic criterion
to mimic the analog-to-digital-converter (ADC) detection threshold (which
is not a fixed energy due to electronic variations that affect the
conversion from deposited energy to a voltage), where
we used the acceptance-rejection method~\cite{NumRec} to sample from
an ansatz\footnote{The chosen functional form for the threshold
function was motivated by the fact that electronic noise
accompanying our energy signals is presumed to be due to a
convolution of many random factors, which, via the central limit
theorem, one would expect to contribute a Gaussian fluctuation to
our energy-signal amplitude. The probability of crossing the ADC
threshold would then be given by the integral of a normal
distribution, i.e. the error function, centered about the threshold
mean. To enable a fit to data, instead a similar function that is
also in the class of sigmoid functions, but with a simpler
functional form, namely the hyperbolic tangent, was chosen instead.} for the likelihood of crossing the ADC threshold, given by the following equation,
\begin{equation}
\label{threshEqn}
\rm{threshold}(E)=0.5\left(1+\tanh\left(\frac{E-c}{d}\right)\right),
\end{equation}
which is plotted in Figure~\ref{ThreshFn}.

\begin{figure}[ht]\begin{center}
\includegraphics[width=1.0\columnwidth,angle=0]{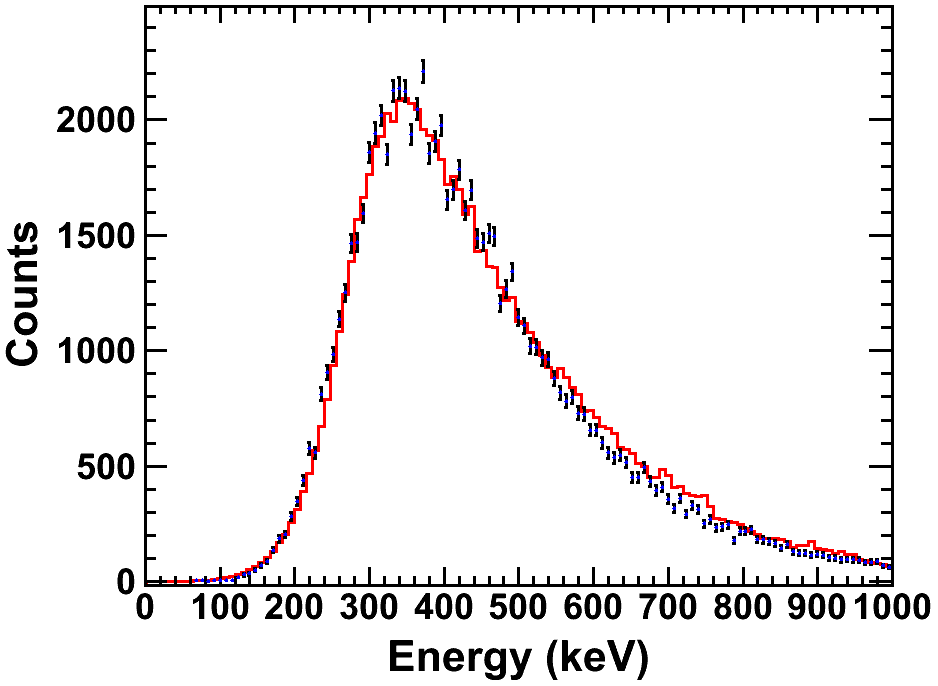}
\caption{(color online.) Comparison of energy deposited by $^{67}$Se
$\beta^{+}$-decays implanted in a 525~$\mu$m-thick DSSD as
simulated by {\tt GEANT4} (solid line) and measured
(points), where a detection
threshold given by Equation~\ref{threshEqn} with $d=60$~keV and
$c=280$~keV has been applied to the
{\tt GEANT4} spectrum. A detector resolution of 150~keV FWHM has
been applied to the simulation data. Note that the simulation and data have the same
binning, where the area of the simulation has been scaled to match
the data.}
\label{Se67spect}
\end{center}\end{figure}

The two free parameters of Equation~\ref{threshEqn} used to mimic
the $\beta$ detection threshold for a $\beta$-particle with energy $E$, namely the
diffuseness $d$ and centroid $c$ of the hyperbolic tangent, were determined
via a $\chi^{2}$-minimization comparison between simulations and
data for $^{67}$Se $\beta$-decays to be 60~keV and 280~keV,
respectively.
Equation~\ref{threshEqn} was applied to
a {\tt GEANT4} simulation of $^{67}$Se $\beta$-decays distributed within a
525~$\mu$m-thick DSSD that mirrored the experimental conditions for the $^{67}$Se
$\beta$-decay measurement discussed in Section~\ref{DataForSimVal}
(Also described in Reference~\cite{DSan14}.).
The simulated location in
the DSSD for the $^{67}$Se $\beta$-decay was sampled from a fit to
data for the implantation planar position and from a depth distribution calculated with
{\tt LISE++}~\cite{Tara08}.
Following the acceptance-rejection method, the $\beta$-energy $E$,
which was sampled in a Monte Carlo fashion from the
$\beta$-energy distribution for $^{67}$Se, was then used as input to
Equation~\ref{threshEqn}. The result of this calculation was
compared to a randomly generated number using a box-like uniform
distribution. This was used to decide whether a $\beta$-particle was
`detected' or discarded. 
The simulation results were in good
agreement with the experimental data, as seen in
Figure~\ref{Se67spect}. We note that the fraction of rejected
events with respect to recorded events for the {\tt GEANT4}
simulation, 66\%, was in agreement with the
experimentally measured ratio of detected $^{67}$Se implants to detected $^{67}$Se
$\beta$-decays, roughly 68\%.

For the DSSD energy-calibration we used a
$^{228}$Th $\alpha$-source, providing energies
$E_{\alpha}=$5.400, 5.685, 6.288, and 6.778~MeV~\cite{Artn97}, and known $\beta$-delayed protons from
$^{20}$Mg, $E_{\rm{p}}=$0.806, 1.679, and
2.692~MeV~\cite{Piec95}\footnote{The first two proton-decay energies
have been measured to higher-precision by~\cite{Wall12}. The results
are consistent with \cite{Piec95} (whose energies were used for this
work and the corresponding work~\cite{DSan14}) within uncertainties and therefore
would not impact the reported results.}, and
$^{23}$Si, $E_{\rm{p}}=$1.32, 2.40, 2.83, and
3.04~MeV~\cite{Blan97}, which were $^{36}$Ar fast-beam fragments measured
within the implantation DSSD. Note that the use of measured energies
from $\beta$-delayed
protons from $^{20}$Mg and $^{23}$Si as calibration points required
the $\beta$-summing correction (discussed in
Section~\ref{Conclusions}) to be applied.
The uncertainty of the proton energies from
$\beta$p emitters $^{20}$Mg and $^{23}$Si, which ranged from
15-60~keV, included a systematic uncertainty associated with the
$\beta$-summing present in published studies~\cite{Piec95,Blan97}
that determined the proton-decay energies. References~\cite{Piec95}
and \cite{Blan97} each corrected for $\beta$-summing, however they
do not elaborate on precisely how this was done.

\subsection{Sensitivity to decay $Q$-value}
\label{Qsens}
Though the reported $\beta$-summing correction technique relies on
knowing the $\beta$-energy spectrum, we demonstrate that our results
are relatively insensitive to uncertainties in the $\beta$-decay $Q$-value, where
here the $Q$-value is to the final state and not necessarily the
ground state.

\begin{figure*}[ht]\begin{center}
\includegraphics[width=1.7\columnwidth,angle=0]{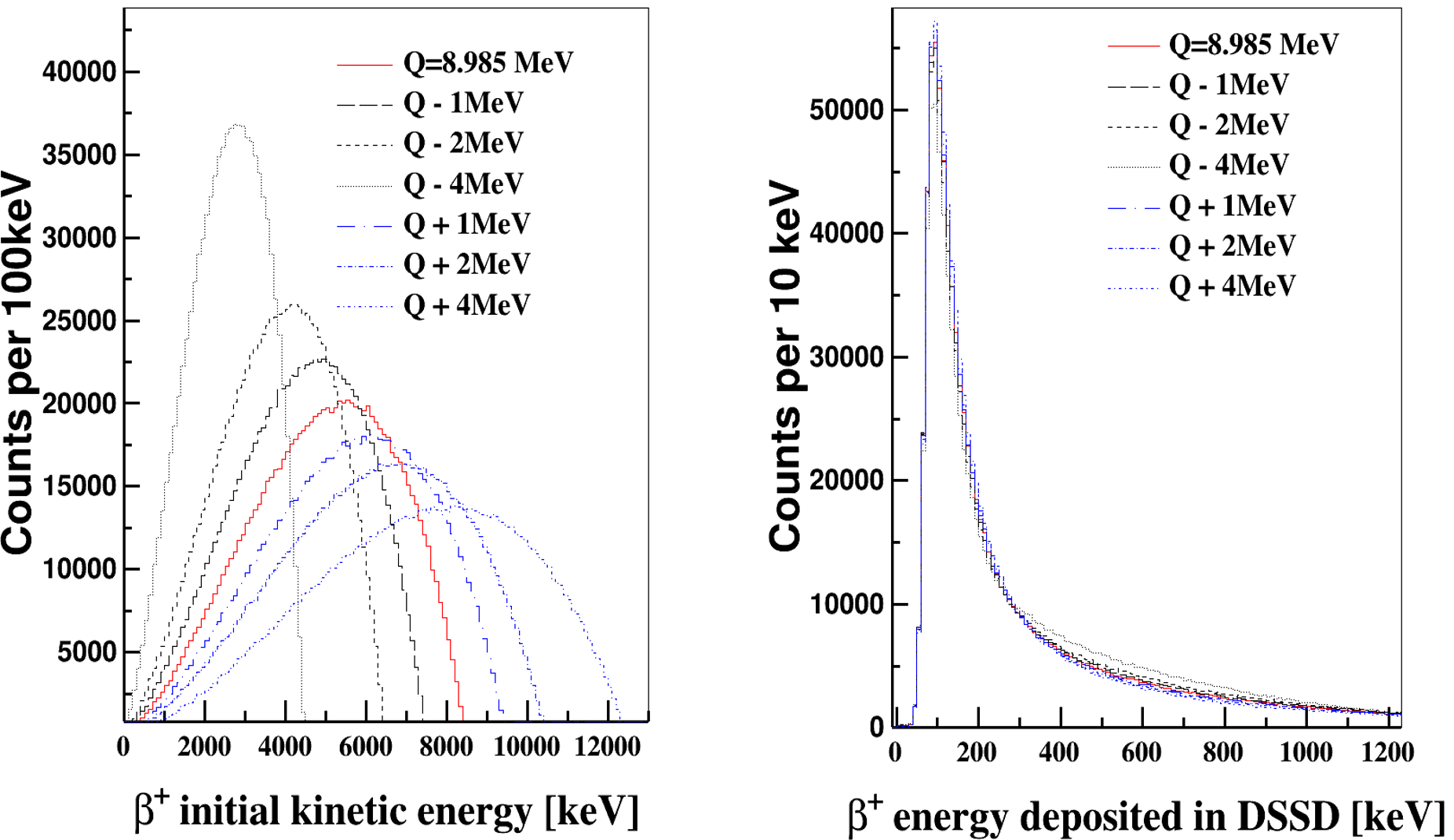}
\caption{(color online.) {\tt GEANT4} simulation results (omitting
detector threshold and resolution effects) for $^{67}$Se
$\beta$-decays, modifying the $Q$-value by up to $\pm$4~MeV from the
nominal value of $\sim$9~MeV. The left panel shows the initial
$\beta$-energy spectra, while the right panel shows the
corresponding $\beta$-energy deposited within the implantation DSSD.}
\label{Qcompare}
\end{center}\end{figure*}

The employed $Q$-values in this work were 9073~keV for
$^{67}$Se~\cite{Audi03} (where a 100\% branch to the ground state
was assumed), 6686~keV for the $^{20}$Mg decay branch to the proton
emitting state
of $^{20}$Na at 806~keV~\cite{Piec95,Audi03},
12083~keV for the $^{23}$Si decay branch to the proton
emitting state of $^{23}$Al
at 2400~keV~\cite{Blan97,Audi03}, 12655~keV for
the $^{69}$Kr decay to the $^{69}$Br ground
state~\cite{DSan14,Audi03}, and 9502~keV for the $^{69}$Kr decay to
the proton emitting state of $^{69}$Br at 2940~keV~\cite{DSan14,Roge14,Audi03}. Since the simulations were
performed for the analysis presented in Reference~\cite{DSan14}, a more
recent mass evaluation has been published~\cite{Audi12}. In no case
does the resultant updated $Q$-value differ from the $Q$-value
employed in this work by more than 90~keV.

\begin{figure}[ht]\begin{center}
\includegraphics[width=1.0\columnwidth,angle=0]{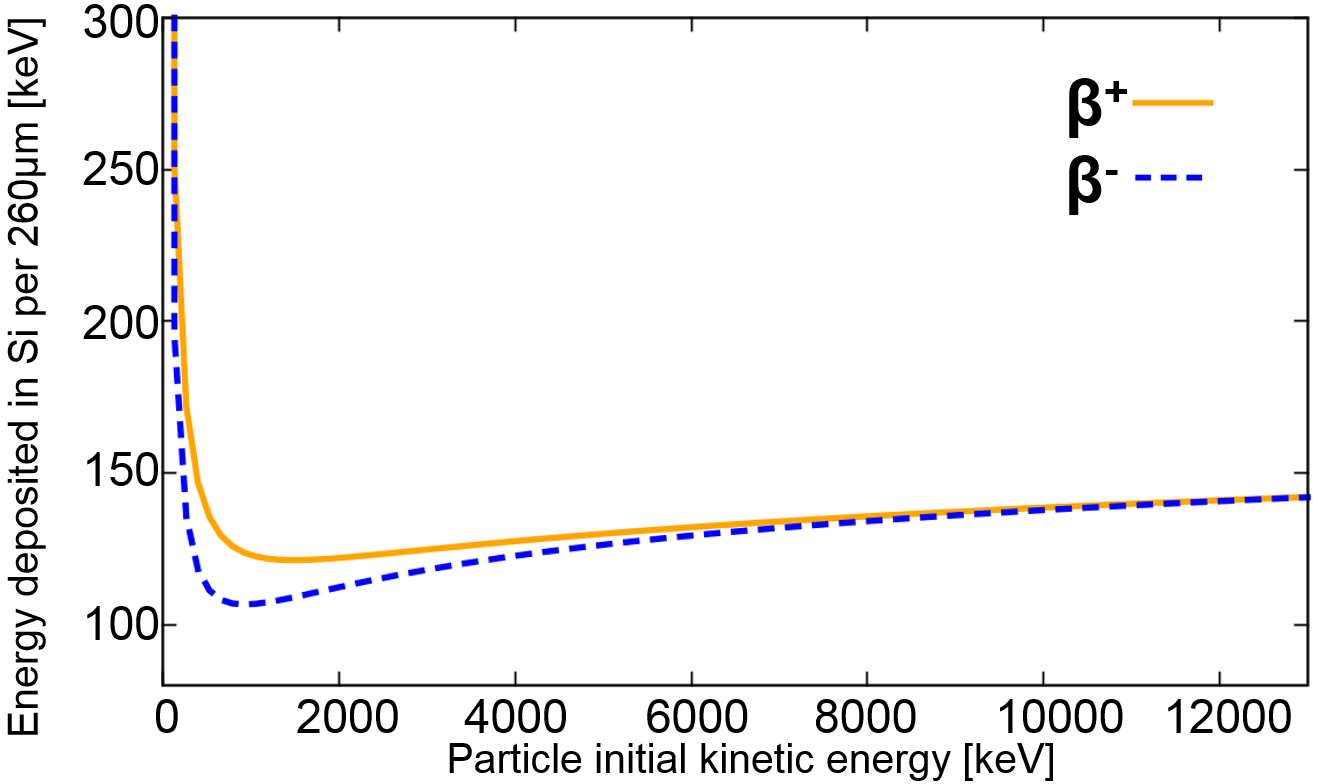}
\caption{(color online.) Analytic energy deposition of $\beta^{+}$
(solid orange line) and $\beta^{-}$ (dashed blue line) particles,
using the relations from Reference~\cite{Rohr54}, in a 260~$\mu$m-thick
layer of silicon.
}
\label{BetaAnalytic}
\end{center}\end{figure}

{\tt GEANT4} simulations were performed for the $^{67}$Se
$\beta$-decay, modifying the $Q$-value from Reference~\cite{Audi12} by
several MeV in order to mimic $\beta$-decay channels into excited states of
the daughter.
Though the initial $\beta$-energy
spectra varied dramatically, the spectra for $\beta$-energy
deposition within the DSSD were remarkably similar for large
modifications to the initial $Q$-value, as shown in
Figure~\ref{Qcompare}. In fact, the $\beta$-energy deposition
spectra were nearly indistinguishable, except for the simulation
which employed a $Q$-value reduced by 4~MeV from the nominal value
of $\sim$9~MeV. This insensitivity to relatively substantial modifications
of the decay $Q$-value is not surprising, but rather expected upon
inspection of the analytic relation for energy deposition of
$\beta$-particles in solid media. Figure~\ref{BetaAnalytic} shows
the relation taken from Reference~\cite{Rohr54} for $\beta$-energy
deposition within 260~$\mu$m of silicon for $\beta$-particles
ranging from 0.1-13~MeV. There, it is apparent that
$\beta$-particles over an energy span of $\sim$2-13~MeV are
expected to deposit nearly the same amount of energy when traveling
through the same thickness of silicon. Therefore, sensitivity to the
$Q$-value, due to the choice of decay final state or the state's
energy uncertainty, is not expected for $Q$-values which result in a
mean initial $\beta$-energy over $\sim$2~MeV, i.e.
$Q\gtrsim$5~MeV. Since all of the $Q$-values for the decays under
consideration in this study are over this threshold (and in all
cases the $Q$-value uncertainty is 0.5~MeV or less), we do not
expect the choice of $Q$-value for our simulations to impact the
results for the $\beta$-energy deposition and therefore for
$\beta$-summing. This insensitivity to the $Q$-value ultimately
allows a higher-statistics $\beta$-delayed proton emission decay branches to be
used to determine the $\beta$-summing correction for a
lower-statistics decay branch with a roughly similar $Q$-value, as was
done for $^{69}$Kr in Reference~\cite{DSan14}.

\section{Novel implantation depth determination method}
\label{Depth}

The amount of $\beta$-summing which occurs for a given $\beta$p
emission depends sensitively on the depth within the DSSD at which the $\beta$p
emission occurs. A deeper depth within the DSSD implies that the
emitted $\beta$ travels through more material while escaping the
implantation DSSD and thus deposits
more energy, on average, when compared to a $\beta$p emission at a
shallower depth.  Though the stochastic nature of implantation depth
and $\beta$ energy-loss reduces the strength of the previous
statement, it remains true for an ensemble of $\beta$p emissions.
Thus, the determination of the mean implantation depth of a $\beta$p
emitter substantially reduces the uncertainty associated with
implementing the $\beta$-summing correction for proton-energy
determination. Typically, the two available methods for determining
implantation depth, described in the following section
(Section~\ref{Compare}), are dependent on the energy
calibration of three DSSDs and similar response to $\beta$-particles
of the same energy for two DSSDs. 
Instead, our method, which is outlined in this section, relies on
the energy calibration of only a single DSSD. Note that the 
insensitivity to the
detector threshold is true for the majority
of $\beta$-delayed proton emission experiments since proton energies
are usually $\gtrsim1$~MeV, while the detection threshold plays no
role above $\sim$0.5~MeV, as seen in Figure~\ref{ThreshFn}.
Consequently, our method provides an efficient way to minimize
systematic uncertainties present for the $\beta$-summing correction.

\begin{figure}[ht]\begin{center}
\includegraphics[width=1.0\columnwidth,angle=0]{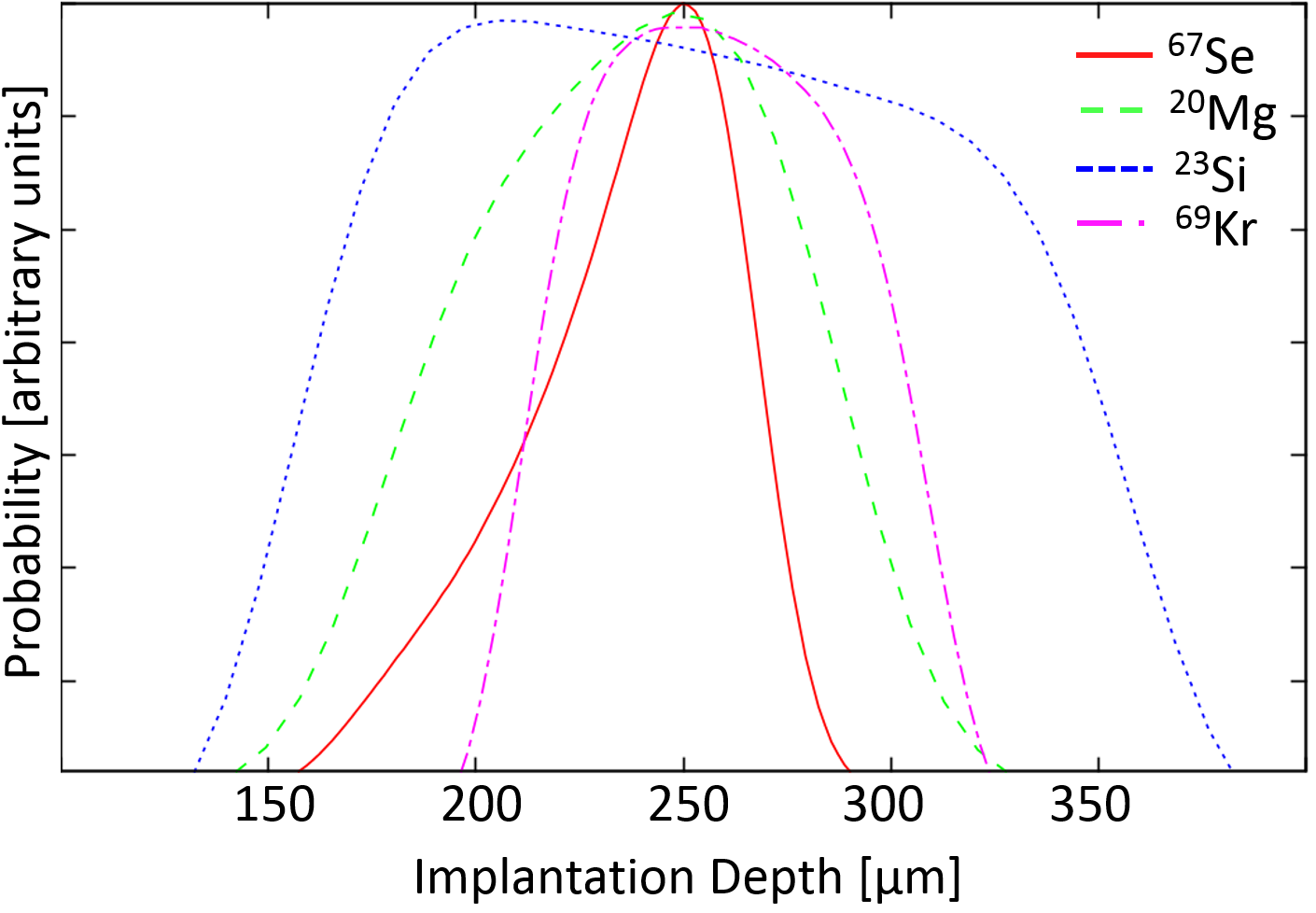}
\caption{(color online.) Probability distributions for the implantation
depths of $^{67}$Se, $^{20}$Mg, $^{23}$Si, and $^{69}$Kr ions within a
525~$\mu$m-thick DSSD calculated
with {\tt LISE++}~\cite{Tara08} using their Straggling Method 1.  The amplitudes and centroids of each distribution have been shifted
arbitrarily for comparison. The distribution shape is primarily
related to the ion's momentum distribution, which depends on the
ion production mechanism, e.g. target thickness and
momentum-acceptance of the fragment separator through which the ion
passes.}
\label{DepthFromLISE}
\end{center}\end{figure}

\begin{figure}[ht]\begin{center}
\includegraphics[width=1.0\columnwidth,angle=0]{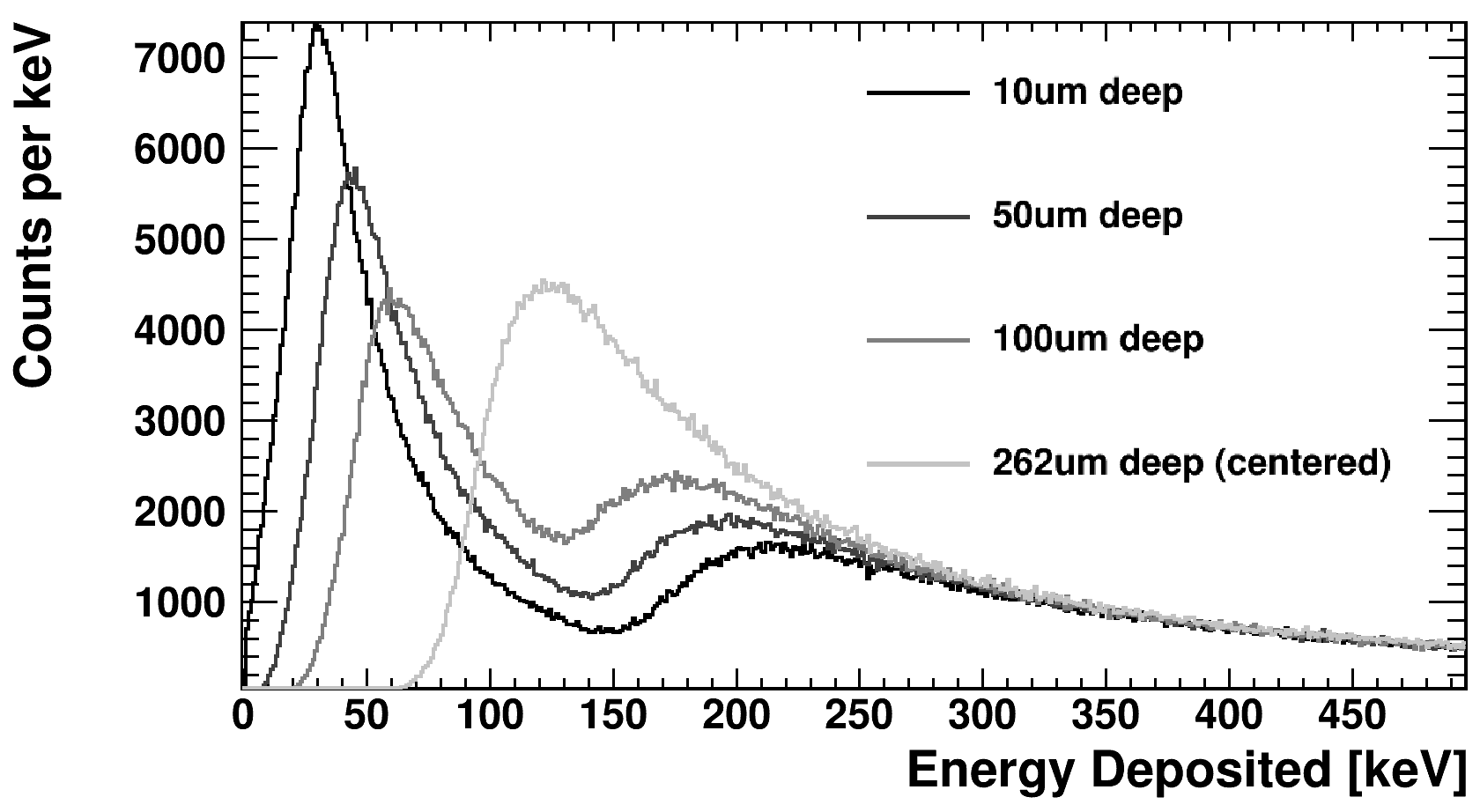}
\caption{ $\beta$-particle energy-deposition histograms from {\tt GEANT4} simulations of
$^{67}$Se $\beta^{+}$-emission at various depths within a 
525~$\mu$m
DSSD, where the detection threshold and detector resolution has not been applied. In each case
the angular distribution of the $\beta^{+}$-emission was isotropic.
Note that all simulated events originated from a single depth
(indicated in the legend) and
were not sampled from a depth distribution.}
\label{DSSDhemiSpect}
\end{center}\end{figure}

In simulating $\beta$-delayed proton emission, we include the
energy and angle of the decay products and the position of the
$\beta$p-emitter within the DSSD, each of which is sampled from a
probability distribution event-by-event.  We sample the positron
energy from the $\beta$-decay distribution given by the
Fermi theory of $\beta$-decay~\cite{Ferm34}.
Corrections for detector threshold and resolution are applied to the discrete proton
energy and deposited $\beta$ energy.  The distributions for proton and positron emission
angles are isotropic, since the orientations of nuclei stopped within
the DSSD are random.  The $\beta$-decay position on the plane
of the DSSD that is perpendicular to the beam-direction
is selected from a two-dimensional skew Gaussian which was fitted to
the measured implantation distribution, which was possible due to
the segmentation of the DSSD into virtual pixels.  As the implantation depth
for a given $\beta$-delayed proton emission event is {\it a priori}
unknown, due to the lack of segmentation in the depth-dimension of
the DSSD, the depth of the $\beta$p-emitter within the
implantation DSSD is selected from a skew Gaussian distribution that
is fitted to the implantation depth distribution
simulated with the multi-purpose simulation tool {\tt
LISE++}~\cite{Tara08} (using their Straggling Method 1) for the $\beta$-delayed proton
emission experiments discussed here. We stress that {\tt LISE++}
is not able to accurately predict the absolute implantation depth of
$\beta$p nuclei within the implantation DSSD, hence the need for
the adjustable aluminum degrader upstream of the BCS (See
Figure~\ref{BCSdraw}.), though we find it accurately predicts the
relative separation in mean implantation depth for ions measured in
the same projectile fragmentation experiment (See
Section~\ref{Compare}.).
The implantation depth distributions for the simulated $\beta$p
nuclei, shown in Figure~\ref{DepthFromLISE}, in general span roughly
\nicefrac{1}{3} of the 525~$\mu$m DSSD thickness. The widths
of the distributions are due to the narrow momentum acceptance of
the fragment separator,
$\pm$0.07\% and $\pm$0.5\% for the $^{78}$Kr and $^{36}$Ar primary
beams,
respectively.  The centroid of an implantation
depth distribution is referred to here as the mean implantation
depth.  Figure \ref{DSSDhemiSpect} shows the impact 
of choosing different discrete depths (not sampling from a depth distribution)
within the DSSD for $\beta$-delayed proton-emission
events on energy deposition of $\beta$-particles emitted within the implantation
DSSD.

The centroid of the depth distribution function from which the
$\beta$p-implantation depth was sampled was changed in
5~$\mu$m
steps from 0 to
260~$\mu$m
between simulations, where
these depths correspond to the upstream-face and center-plane of the
DSSD, respectively.  Due to
the symmetry of the detector, equivalent results are obtained by
choosing centroids from 525 to
265~$\mu$m.
However, this degeneracy
is not an issue, as an accurate $\beta$-summing correction only
relies on knowing the distance between the centroid and closest DSSD
planar surface.  As will be shown in the following section, we find
our mean implantation depth determination method has a precision on the order
of tens of microns, and thus we find a
5~$\mu$m
grid to be sufficient.

\begin{figure}[ht]\begin{center}
\includegraphics[width=1.0\columnwidth,angle=0]{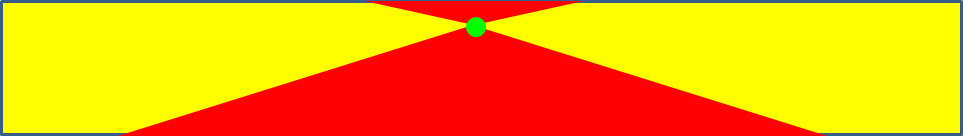}
\caption{(color online.) Cartoon representation of a hypothetical
$\beta$-particle emission (green circle) that occurs within a fixed
solid angle (red areas) towards opposing DSSD (yellow rectangle) hemispheres. The relative average amount of
material through which a $\beta$-particle travels changes depending on the
location of the $\beta$-particle emission, thus affecting the
energy-deposition
spectrum, as seen in Figure~\ref{DSSDhemiSpect}. }
\label{DSSDhemi}
\end{center}\end{figure}

We find the profile of the histogram of energy deposition within
the implantation DSSD is sensitive to the mean implantation depth, as demonstrated in Figure \ref{DSSDhemiSpect}.
Qualitatively, this can be understood by considering a
single depth for $\beta^{+}$-emission, allowing a positron to be
emitted within a given solid angle toward the downstream or upstream
hemisphere of the DSSD, where the central plane of the DSSD defines the
boundary between hemispheres, as in Figure~\ref{DSSDhemi}.  We
justify this approximation made for the
purpose of demonstration given the peaked nature of the $\beta$p-implantation depth distribution ($\rm{FWHM}\sim0.15$~mm) and the narrow thickness of the
DSSD (0.525~mm) in comparison to its length and width
($40\times40$~mm$^{2}$).  From
Figure~\ref{DSSDhemi} it is apparent that, by
choosing a depth which is off-center, much more silicon is
contained within a given solid angle for the thicker hemisphere as
opposed to the thinner hemisphere.  As energy deposition is linearly
related to the length of detector traversed for a
particle whose energy remains nearly constant, a condition which the positron roughly meets, it is
apparent that the mean energy deposited will be higher for the
events traversing the thicker hemisphere.  For this simplified
situation we obtain a low-energy peak for $\beta$-events in the hemisphere
directed toward less silicon (closer to the surface) and a
high-energy peak for $\beta$-events in
the hemisphere directed toward more silicon (further from the
surface).
By considering the histograms obtained
from each hemisphere together, we obtain double-peaked
histograms like those shown in Figure~\ref{DSSDhemiSpect}.
The high and low energy peaks are blended together by
including the depth and $\beta$-energy distributions into the
simulation, however the general effect remains.

\begin{figure}[ht]\begin{center}
\includegraphics[width=1.0\columnwidth,angle=0]{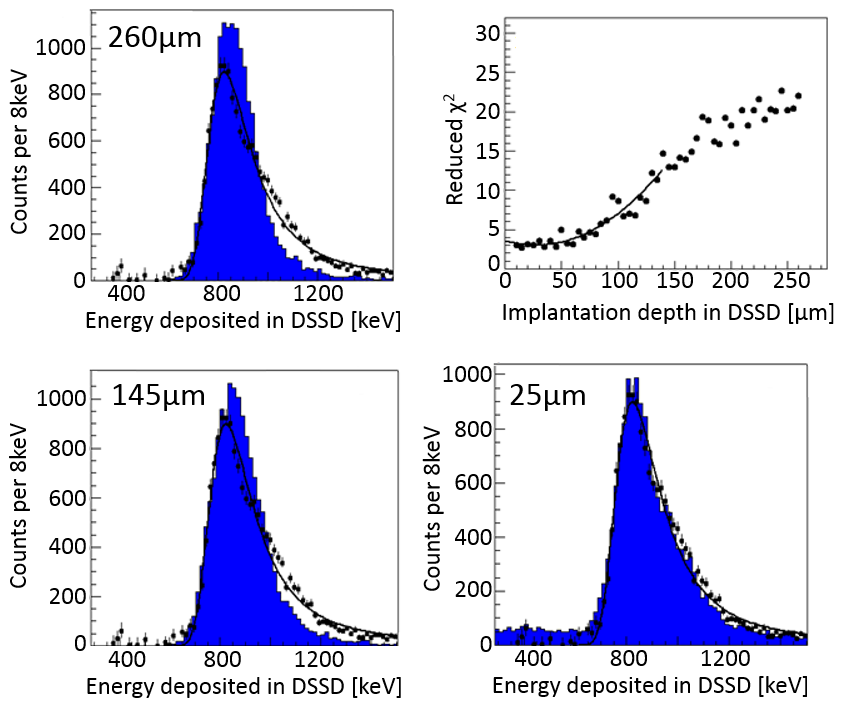}
\caption{(color online.) Comparison between {\tt GEANT4} simulations
(filled histogram) of $^{20}$Mg total (proton + recoil + $\beta$) energy
deposition corresponding to the 806~keV proton decay within a
525~$\mu$m-thick DSSD and
experimental data (points) for simulated mean implantation depths of
260, 145, and 25~$\mu$m
from the upstream planar
surface of the implantation DSSD. The simulation histogram
area has been scaled to match the data and a fit to data with a Landau
distribution (black line) is included for reference.
The upper-right panel contains the reduced-$\chi^{2}$ between
simulated total energy-deposition histograms and experimental spectra as a
function of the $\beta$p nucleus mean implantation depth, including
the quadratic fit.}
\label{Mg20sim}
\end{center}\end{figure}

\begin{figure}[ht]\begin{center}
\includegraphics[width=1.0\columnwidth,angle=0]{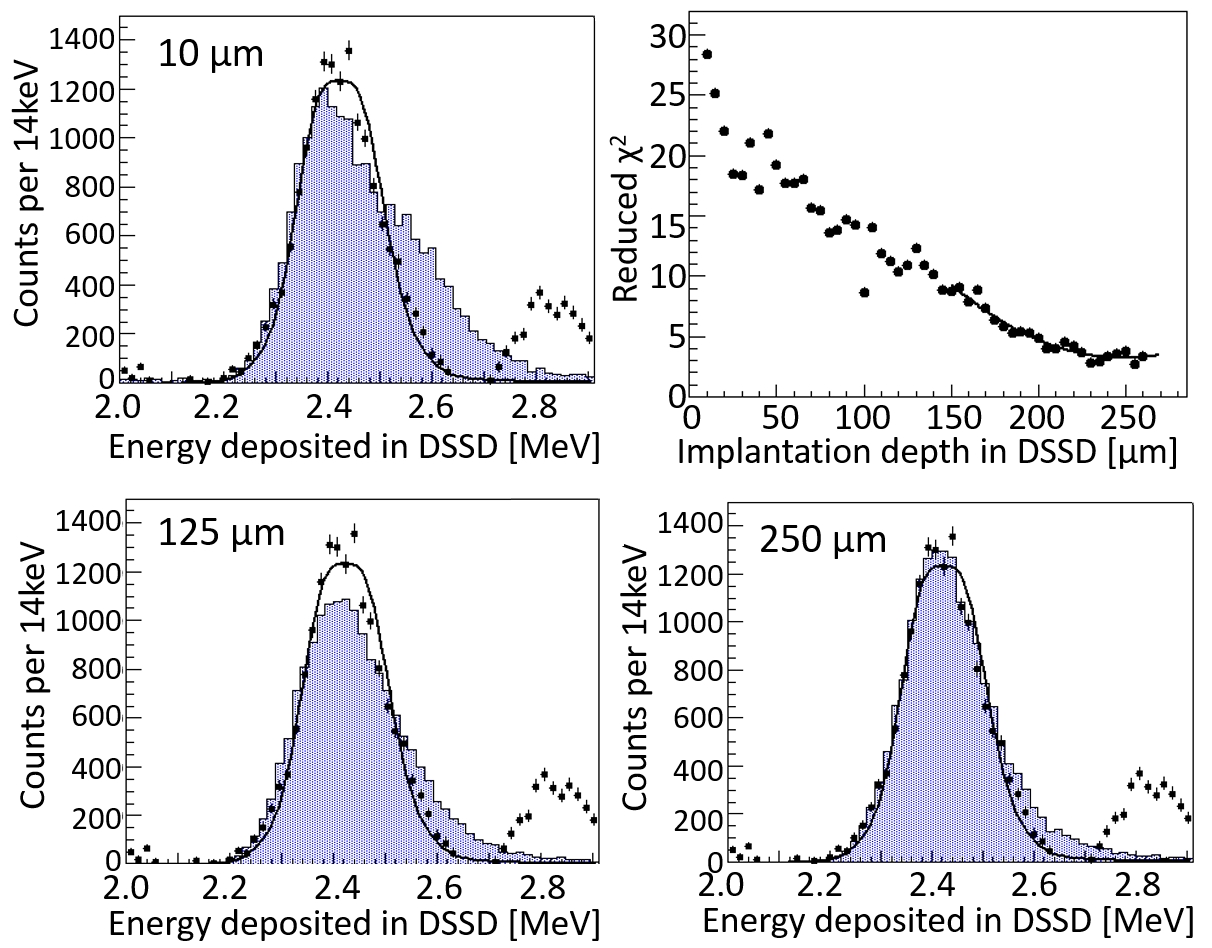}
\caption{(color online.) Same as Figure~\ref{Mg20sim}, but for
$^{23}$Si $\beta$p events at mean implantation depths of 10, 125,
and 250~$\mu$m. Note that the higher-energy peak in the data is from
a separate proton-emitting branch.}
\label{Si23sim}
\end{center}\end{figure}

To test our depth determination method, we use $\beta$-delayed proton emission of $^{20}$Mg and $^{23}$Si,
each of which were produced via projectile fragmentation of
$^{36}$Ar and implanted in a 525~$\mu$m-thick
DSSD\footnote{$\beta$-decay data were not used to assess the impact
of implantation depth on $\beta$-summing
since those data are far more sensitive to the DSSD detection threshold}.
For the proton-decay energy from each source with the highest
statistics, $E_{\rm{p}}=0.806$~MeV for $^{20}$Mg and
$E_{\rm{p}}=2.40$~MeV for $^{23}$Si, we perform our simulation
for the aforementioned range of $\beta$p-implantation depth
centroids and distributions.  A reduced-$\chi^{2}$ value is
calculated for each mean implantation
depth by comparing the simulated total (proton + recoil + $\beta$) energy
deposition distribution to the data.  The minimum of the fit to
reduced-$\chi^{2}$ results is taken as the mean implantation depth,
whereas the uncertainty is the difference between this depth and the
depth for which the fit to reduced-$\chi^{2}$ results is $+1$
greater than the minimum. Figure~\ref{Mg20sim} shows sample spectra
comparing the total energy-deposition histogram of the 806~keV
proton decay from
$^{20}$Mg to simulations using three different mean implantation
depths. In this way we are able to assign a mean implantation depth for both
$^{20}$Mg and $^{23}$Si (See Figure~\ref{Si23sim}.), which we compare to other depth
determination methods in the following section.
The precise mean implantation depth is important because it
corresponds to a reduction in the proton-decay energy uncertainty that
results from the $\beta$-summing correction, as discussed in
Section~\ref{Conclusions}.  This method provides a way to
determine the mean implantation depth of a $\beta$p emitter which can
be used in addition to or in lieu of  (if they are not feasible) the
other methods of mean implantation depth determination discussed in the following section.

\section{Other mean implantation depth determination methods}
\label{Compare}
In order to validate our mean implantation depth determination
method, `Method 1', we compare it to other available methods. 
In addition to the method described in the prior section, we use two independent methods to determine the mean implantation
depth for the $\beta$-delayed proton emitters $^{20}$Mg and $^{23}$Si.

\subsection{Method 2: Implantations per implantation DSSD and upstream or downstream DSSDs}
For method 2 we look at the fraction of implantations of a
given nucleus that occurs in the DSSDs which are neighbors to the
implantation DSSD.  This method requires that a nucleus is poorly
centered within the central (implantation) DSSD in the BCS detector stack, and requires consideration of the asymmetry of the nucleus's implantation depth distribution.
Integration is performed over the implantation depth probability
distribution over the longitudinal range
encompassing each DSSD until the fraction of events occurring within
each detector matches the fraction of implantation events observed
for that nucleus in each DSSD. The uncertainty quoted for this
		  method is the simulated mean implantation depth range for
		  which the ratio of events deposited in the neighboring and
		  implantation DSSDs agrees with the measured ratio within its
		  statistical uncertainties.

This method is naturally more sensitive for nuclei whose mean implantation depth is
close to the surface of the implantation DSSD, provided that nucleus
has a narrow implantation depth distribution, and less sensitive (or
impossible) for a nucleus well centered within the implantation DSSD.  Since calculations of $^{20}$Mg implantation with
{\tt LISE++}~\cite{Tara08} result in a depth distribution with
$\rm{FWHM}$
150~$\mu$m,
whereas the DSSD is
525~$\mu$m-thick, this
method works well. However, it is apparent that this implantation
depth method is reliant on an accurate prediction of the
implantation depth distribution shape.  

\begin{figure*}[ht]\begin{center}
\includegraphics[width=2.0\columnwidth,angle=0]{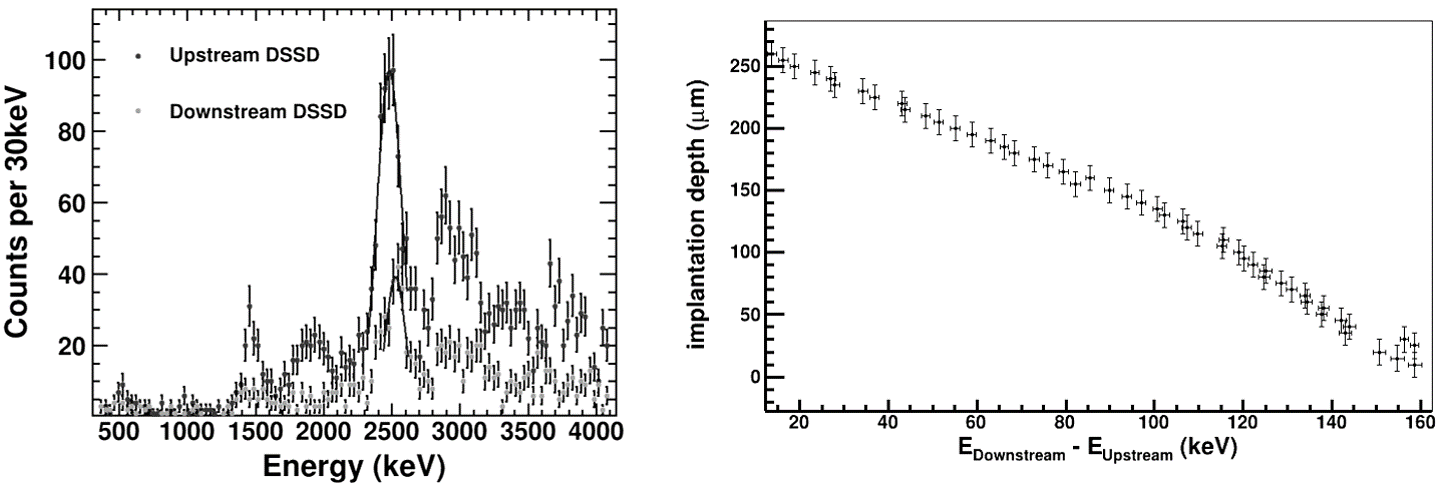}
\caption{(color online.) Mean implantation depth determination
`Method 3' for $^{23}$Si which used
the difference in mean energy deposition within DSSDs
upstream (dark-gray points) and downstream (light-gray points) of the
implantation DSSD, where the blue lines are Gaussian fits (left
panel) to $\beta$-energy deposition peaks associated with
$\beta$-delayed proton emission of $^{23}$Si through the
proton-emission channel ($E_{\rm{p}}=2.40$~MeV) with the highest statistics. The left panel shows the energy distributions,
whereas the right panel shows the energy difference between means of
the upstream and downstream DSSD $\beta$-energy deposition distributions in relation to
the implantation depth within the implantation DSSD for {\tt GEANT4}
simulations. The measured mean-energy difference, $\langle
E_{\rm{DSSD1}}\rangle-\langle E_{\rm{DSSD3}}\rangle=45\pm9$~keV,
corresponds to an implantation depth of
215$\pm$14~$\mu$m.
}
\label{Si23diffSpect}
\end{center}\end{figure*}

\subsection{Method 3: Mean $\beta$-energy difference between upstream and
downstream DSSDs}

For method 3, we compare the separation in mean energy deposited
by positrons in the DSSDs which are upstream and downstream of the
implantation DSSD (DSSD2 in Table~\ref{DetectorTable}), selecting only decays within a small solid angle
about the axis perpendicular to the DSSD planes that correspond to
the proton-emission channel of interest.  Positrons which
traveled
through more silicon in the implantation DSSD on average lost
more energy in the implantation DSSD than positrons which traveled
through less silicon. 
Therefore, a slightly different $\beta$-energy deposition is
expected between the DSSDs neighboring the implantation DSSD, since
they
are separated from the mean implantation depth by different amounts
of silicon.
This phenomenon
was
used by Reference~\cite{Piec95} to gate on events that had minimal
$\beta$-summing in their implantation DSSD. 

Simulations were
employed to map between mean implantation depth and the mean
$\beta$-energy deposition
difference between $\beta$-energy distributions for the upstream and downstream DSSDs. The results of
this process for $^{23}$Si are shown in Figure~\ref{Si23diffSpect},
where the reported mean implantation depth uncertainty is the
simulated depth range that resulted in an upstream-downstream mean
$\beta$-energy difference
within the range given by the fitted mean-energy uncertainties
summed in quadrature.
We note that this method of mean implantation depth determination
relies upon the DSSDs which are upstream and downstream of the
implantation DSSD having a similar response to $\beta$-particles of
the same energy.  

\subsection{Comparison between mean implantation depth determination
methods}
We find agreement
between each of the three mean implantation depth determination
methods for $^{20}$Mg and $^{23}$Si, as seen in
Table~\ref{DepthTable}. The relative separation in mean implantation
depth between $^{20}$Mg and $^{23}$Si is in good agreement with the
separation predicted by {\tt LISE++}, 
180~$\mu$m. Here we stress that the advantage of the
method presented in Section~\ref{Depth} (Method 1 of
Table~\ref{DepthTable}), is that it only requires a single DSSD,
namely the implantation DSSD, and therefore is only
sensitive to the energy calibration and electronic noise of one
rather than three DSSDs and it is generally not sensitive to the
detector threshold of any DSSD. Additionally, we note that Methods 1 and 3
are more precise for $\beta$p nuclei deposited closer to the center of the
implantation DSSD, while Method 2 excels
  for $\beta$p nuclei deposited nearer to the implantation DSSD
			 surface.

\begin{center}
\begin{table}
  \caption{Mean implantation depth within a 525~$\mu$m-thick DSSD as determined by energy
  deposition profile $\chi^{2}$-minimization, implantations
  per DSSD, and mean-energy difference between upstream and downstream
  DSSDs,
  which are termed Methods 1, 2, and 3, respectively. $^{23}$Si has
  no value for Method 2 since all $^{23}$Si nuclei were implanted
  within the central DSSD, i.e. DSSD2 in Table~\ref{DetectorTable}.
  It is apparent that Methods 1 and 3 are more precise for
  centrally-deposited $\beta$p nuclei, while Method 2 is preferred
  for $\beta$p nuclei deposited nearer to the surface of the
			 implantation DSSD.}
  \centering
  \begin{tabular}{ c  c  c  c }
  \hline
  $\beta$p emitter & Method 1 &   Method 2 &   Method 3 \\ \hline
  $^{20}$Mg        & 27(36)~$\mu$m & 27(6)~$\mu$m & 61(50)~$\mu$m \\ 
  $^{23}$Si        & 250(40)~$\mu$m & --- & 215(14)~$\mu$m \\ 
  \end{tabular}
  \label{DepthTable}
  \end{table}
\end{center}

\section{Determination of the $\beta$-summing correction}
\label{Conclusions}

The accurate mean
implantation depth determination methods described in the previous
sections (Sections~\ref{Depth} and~\ref{Compare}) enable an accurate
determination of the average $\beta$-energy deposited in $\beta$-delayed
proton emission events within a DSSD, and thus allow for an accurate
proton-decay energy to be determined. The methods' typical precision of
tens of microns results in an uncertainty in the $\beta$-summing
correction which is on the order of tens of keV for the cases
presented here. 

\begin{figure*}[ht]\begin{center}
\includegraphics[width=2.0\columnwidth,angle=0]{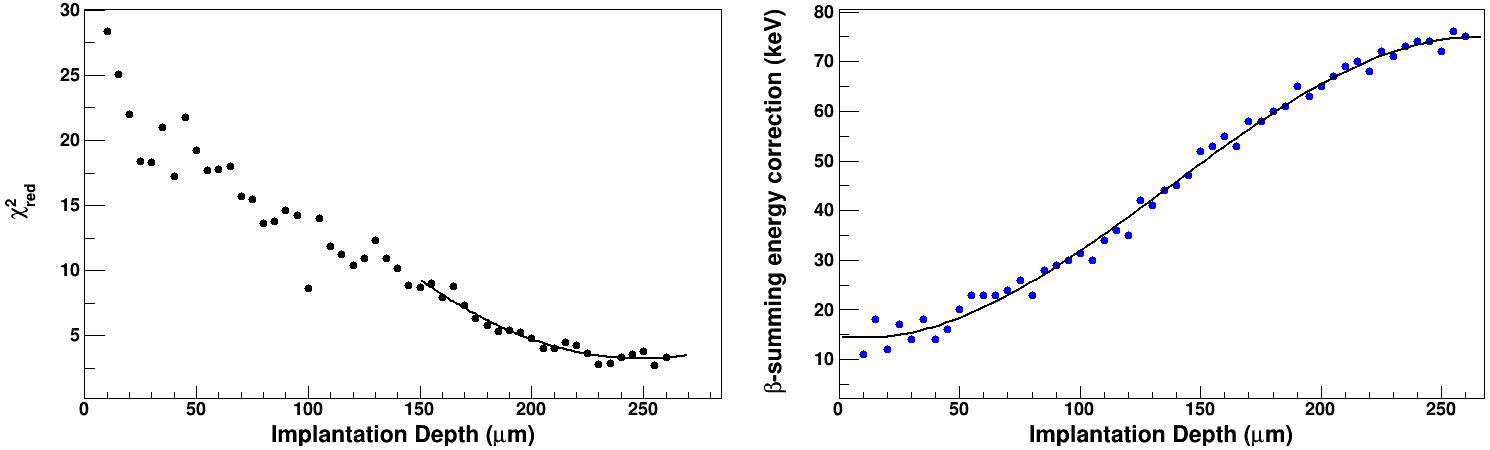}
\caption{(color online.) Reduced-$\chi^{2}$ minimization from mean implantation
depth determination Method 1 for $^{23}$Si as a function of mean implantation
depth (left panel), including a quadratic fit, demonstrating a mean implantation depth of
250$\pm$40~$\mu$m. The right panel shows the corresponding
$\beta$-summing correction applied to $^{23}$Si proton energy
deposition peaks as a function of mean implantation depth, where a
third-order polynomial fit is included to guide the eye.}
\label{Si23summingCorrection}
\end{center}\end{figure*}

The relationship between $\beta$-summing
and mean implantation depth within a 
525~$\mu$m DSSD for $^{23}$Si is
shown in Figure~\ref{Si23summingCorrection}.
Under our conditions, an implantation depth distribution with
$\sim$150~$\mu$m
$\rm{FWHM}$ in a
525~$\mu$m-thick DSSD, we are able to
determine the mean implantation depth within
$\sim$40~$\mu$m;
however,
more accurate results would be possible for a narrower depth distribution
in a thinner detector.  As seen in
Figure~\ref{Si23summingCorrection}, the mean implantation depth
uncertainty of 
40~$\mu$m
translates into a $\beta$-summing correction
uncertainty of anywhere from 5 to 25 keV, depending on which place
along the depth--correction relationship the mean implantation depth
is located. Similarly for $^{20}$Mg, whose $\beta$-summing correction was
found to be between 10 and 85~keV depending on the mean implantation
depth,
the $\beta$-summing correction uncertainty was between 5 and 15 keV for the
range of possible mean implantation depths.
Given the determined
implantation depth, the actual $\beta$-summing correction for the
deduced mean implantation depth was 18$\pm$6~keV.

To correct for $\beta$-summing and determine the proton-decay energy from a $\beta$-delayed proton
emitting nucleus, we incrementally iterate over
proton-decay energies in the Monte Carlo simulation for each mean
implantation depth. $\chi^{2}$ minimization between the energy-deposition histograms, which include the energy deposited by the
proton+recoil and $\beta$-summing, from the data and
simulations simultaneously results in a proton-decay energy and mean
implantation depth.  This procedure has been used for the $\beta$-delayed
proton emission of $^{69}$Kr (using the 2.94~MeV proton-decay peak) to obtain a summing correction of
79$\pm$12~keV\cite{DSan14}.

Therefore, the full process of applying the $\beta$-summing
correction to obtain a precise proton energy from a $\beta$-delayed
proton emission measurement in a DSSD is the following:
\begin{enumerate}
\item Obtain the $\beta$p nucleus surface implantation distribution
by fitting to the implantation distribution measured with the
pixelated DSSD.
\item Obtain the $\beta$p implantation depth distribution by
fitting to results from {\tt LISE++} simulations that mimic
experimental conditions, i.e. ion production, transport, and
detector materials.
\item Perform Monte Carlo simulations for a given $\beta$-delayed
proton decay branch with {\tt GEANT4}, randomly
selecting the $\beta$ energy and angle, proton angle, and $\beta$p
decay location, for mean implantation depths ranging from the
detector surface to detector center.
\item Perform $\chi^{2}$-minimization between the simulated and
measured total (proton + recoil + $\beta$) energy-deposition histograms,
allowing the location of the histogram peak to vary.
\item For the simulation which yields the minimum
reduced-$\chi^{2}$, the location of the histogram peak simultaneously yields the
proton-decay energy and the amount of $\beta$-summing (since the
proton-decay
energy deposited in the simulation can be plotted simultaneously
with the total energy-deposition).
\end{enumerate}

\section{Conclusions}

In summary, we present an approach to address the problem of
$\beta$-summing in the measurement of proton-decay energies of
$\beta$-delayed proton-emitting nuclei detected via implantation
within a DSSD. We demonstrate that determination of the
mean implantation depth of the $\beta$p nucleus within the
implantation DSSD subsequently determines the magnitude of
$\beta$-summing. We describe three methods to determine
the mean implantation depth, two of which require DSSDs located
upstream and downstream of the implantation DSSD and a third which
depends only on the total (proton + recoil + $\beta$) energy-deposition histogram of the implantation
DSSD and is generally insensitive to the detector threshold
uncertainty. For the cases discussed, these techniques are capable of determining the mean
implantation depth of a $\sim100$~MeV/u $\beta$p nucleus within a
$\sim$0.5~mm-thick DSSD to within tens of microns, corresponding to
a $\beta$-summing correction uncertainty of $<$25~keV.

\section{Acknowledgements}
\label{Acknowledgements}
This material is based upon work supported by the National Science
Foundation under Grants Nos. PHY-0822648 and PHY-1430152.




\bibliographystyle{model1a-num-names}
\bibliography{BetaSummingCorrection}

\end{document}